\RequirePackage{lineno}

\documentclass[preprint,aps,prl,showpacs,superscriptaddress,floatfix]{revtex4}

\usepackage{epsfig}
\usepackage{color}
\newcommand {\op}	{\rho\sigma L}
\newcommand {\Ncoll}	{N_{\rm coll}}

\begin{document}
\title{Elliptic flow from Coulomb interaction and low density elastic scattering}
\author{Yuliang Sun}
\affiliation{School of Science, Huzhou University, Huzhou, Zhejiang 313000, P.R.~China}
\author{Qingfeng Li}
\affiliation{School of Science, Huzhou University, Huzhou, Zhejiang 313000, P.R.~China}
\affiliation{Institute of Modern Physics, Chinese Academy of Sciences, Lanzhou, Gansu 730000, P.R.~China}
\author{Fuqiang Wang}
\email{fqwang@zjhu.edu.cn}
\affiliation{School of Science, Huzhou University, Huzhou, Zhejiang 313000, P.R.~China}
\affiliation{Department of Physics and Astronomy, Purdue University, West Lafayette, Indiana 47907, USA}

\date{\today}

\begin{abstract}
In high energy heavy ion collisions and interacting cold atom systems, large elliptic flow anisotropies have been observed. For the large opacity ($\op\sim 10^{3}$) of the latter hydrodynamics is a natural consequence, but for the small opacity ($\op\sim 1$) of the former hydrodynamic description is questionable. To shed light onto the situation, we simulate the expansion of a low density Argon ion (or atom) system, initially trapped in an elliptical region, under the Coulomb interaction (or elastic scattering). Significant elliptic anisotropy is found in both cases, and the anisotropy depends on the initial spatial eccentricity and the density of the system. The results may provide insights into the physics of anisotropic flow in high energy heavy ion collisions and its role in the study of quantum chromodynamics.
\end{abstract}


\maketitle

\section{Introduction}
In high energy heavy ion collisions~\cite{Gyulassy:2004zy} and interacting cold atom systems~\cite{O'Hara:2002zz}, large elliptic flow anisotropies have been observed. The observations can be described by hydrodynamics~\cite{Heinz:2013th,Schaefer:2014awa} with low viscosity to entropy density ratio, close to the conjectured quantum low limit of $1/4\pi$~\cite{Kovtun:2004de}. This suggests that both these systems are strongly interacting~\cite{Gyulassy:2004zy,Heinz:2013th,Schaefer:2014awa,Ollitrault:1992bk,Shuryak:2008eq,Schafer:2009dj}, because it is generally believed that strong interactions are a precursor for large anisotropic flow. 

Interaction strength can generally be quantified by the opacity $\op$, where $\rho$ is the particle density in the system, $\sigma$ is the interaction cross section between the particles, and $L$ is the system size. The opacity is the inverse of the Knudsen number where $1/\rho\sigma$ is the mean free path of particles in the system.
The opacity of the cold atom system, $\op\sim 10^3$~\cite{O'Hara:2002zz}, is indeed large and hydrodynamics is a natural consequence~\cite{Thomas2009665c}. A recent parton transport study~\cite{He:2015hfa} by AMPT (A Multi-Phase Transport) and MPC (Molnar's Parton Cascade) indicates, however, that the opacity of heavy ion collisions is small, $\op\sim 1$. For such a small opacity (i.e.~weak coupling), the anisotropic flow should be small as shown by low-density limit studies~\cite{Heiselberg:1998es,Voloshin:1999gs}. However, anisotropic flow from transport models are large, as same as those observed in heavy ion collision data which are described by hydrodynamics. At low opacity, however, hydrodynamic description is questionable~\cite{Bozek:2011if,Huovinen:2008te,Molnar:2009pq}. In fact, the transport model study suggests that the large anisotropic flow is generated by the escape mechanism at low densities~\cite{He:2015hfa}.

The constituents of the two systems and the nature of the interactions are vastly different. Relativistic heavy ion collisions involve the quark and gluon degrees of freedom under the strong interaction governed by quantum chromodynamics (QCD). The cold atom system is composed of neutral atoms under the electromagnetic interaction. That both these systems develop large elliptic anisotropy suggests that the physics is universal, not depending on the nature or details of the interactions. In fact, it is generally believed that the generation of large anisotropic flow is a universal strong coupling physics~\cite{Ollitrault:1992bk}. However, the low opacity suggested by the recent transport model study~\cite{He:2015hfa} suggests that the physics may be universal even including weak coupling. Weak coupling should be relevant in low energy heavy ion collisions, where elliptic flow has been also extensively studied~\cite{Reisdorf:1997fx,Herrmann:1999wu}. There, it is generally established that, besides the strong nuclear force, the Coulomb interaction is also important for the development of collective flow~\cite{Li:2011zzp}.

To shed further lights onto the physics of collective flow development, we simulate the anisotropic expansion of low density Argon ion and atom systems under the Coulomb interaction and short-range contact interaction (elastic scattering), respectively.

\section{Simulation details}
We simulate systems of Argon $^{40}Ar^{+}$ ions and Argon atoms. The particles are initially trapped in an elliptical region, uniformly distributed. To set the half axle lengths of the ellipse, we consider the situation of Au+Au collisions with impact parameter $b=8$~fm, where the transverse overlap area is an ellipsoid with aspect ratio of $a:b=1:2.4$~\cite{Miller:2007ri}. Since we study the systems on atomic length scale, not nuclei, the half axle lengths of the systems are set to $a=10$~nm, $b=24$~nm, and $c=50$~nm. The ellipsoidal eccentricity is quantified by $\varepsilon_2=(b^2-a^2)/(b^2+a^2)$. The aspect ratio is varied in the range from $1:1$ to $1:3.6$ in our study, keeping the volume fixed. The number of particles $N$ is varied from 100 to 5000. The particle momenta are determined by the Boltzmann distribution at temperature $T=300$~Kelvin. The initial average thermal velocity of the particles is 432~m/s.

We release the trap and let the system expand. Two cases of interactions are studied: the Coulomb interaction and the short-range contact interaction (elastic scattering) for the Argon ion and atom system, respectively. The motion of the particles under the respective interactions are calculated using the time-step approach. The time step is set to $10^{-4}$~ns. During each step, particles undergo rectilinear motion. The positions and velocities of all particles in the system are recalculated after each time step. The kinetic energies of the particles in our simulations are typically small, not enough to ionize the atoms or further ionize the ions. So we do not consider ionization in our simulations.

Coulomb interactions have been extensively studied by Coulomb explosion simulations where an incident ion excites the electron cloud in a solid and interacting with the transiently ionized atoms~\cite{PhysRevLett.88.165501,BRINGA20031,SZENES201376}. For our case of the Coulomb interaction, the calculations are significantly simplified.
The net Coulomb force on each ion is calculated from all other ions at their respective positions. The typical Coulomb force for our system of 1000 ions is $10^{-10}$~N. This gives a typical acceleration of $10^{6}$~nm/ns$^2$ for the ions.
The acceleration is used to calculate the new position and velocity after each time step. The typical Coulomb potential energy is on the order of 1~eV, larger than the thermal energy of approximately 0.025~eV, so only the Coulomb force is considered for the ion system.

For the case of the short-range contact interaction, an elastic scattering is considered to happen when two atoms are within a distance of 1~nm (i.e.~cross section $\sigma=\pi$~nm$^2$). Since the atom speed in our simulation of elastic scatterings is typically a few hundred m/s, atom pairs would not enter into a cross section of interaction region and then leave the region within a single step. This ensures that situations do not happen where two particles should really have interacted but the simulation program does not know it. This guided us in the choice of the time step size.
The scattering angle in the rest frame of the two incoming atoms is set to be isotropic (i.e.~chosen randomly in the $4\pi$ solid angle). The velocities of the outgoing atoms after an elastic scattering is determined by energy and momentum conservations. Note that although an atom system is simulated, there is nothing special about atoms, and one could have equally well simulated quarks, neutrons, molecules, or billiard balls with properly scaled dimensions and scattering cross sections.

We investigate how anisotropies develop in the expansion of the systems. The anisotropy is quantified in momentum space by~\cite{Poskanzer:1998yz}
\begin{equation}\label{1}
v_{2}=\langle \cos(2\phi)\rangle,
\end{equation}
where $\phi$ is the azimuthal angle (on the $x$-$y$ plane) relative to the short-axis ($x$) of the trap. The anisotropy magnitude is studied as functions of the initial trap aspect ratio and the particle density of the system.

\section{Results and discussions}
\begin{figure}[hbt]
\centering
\includegraphics[width=0.75\textwidth]{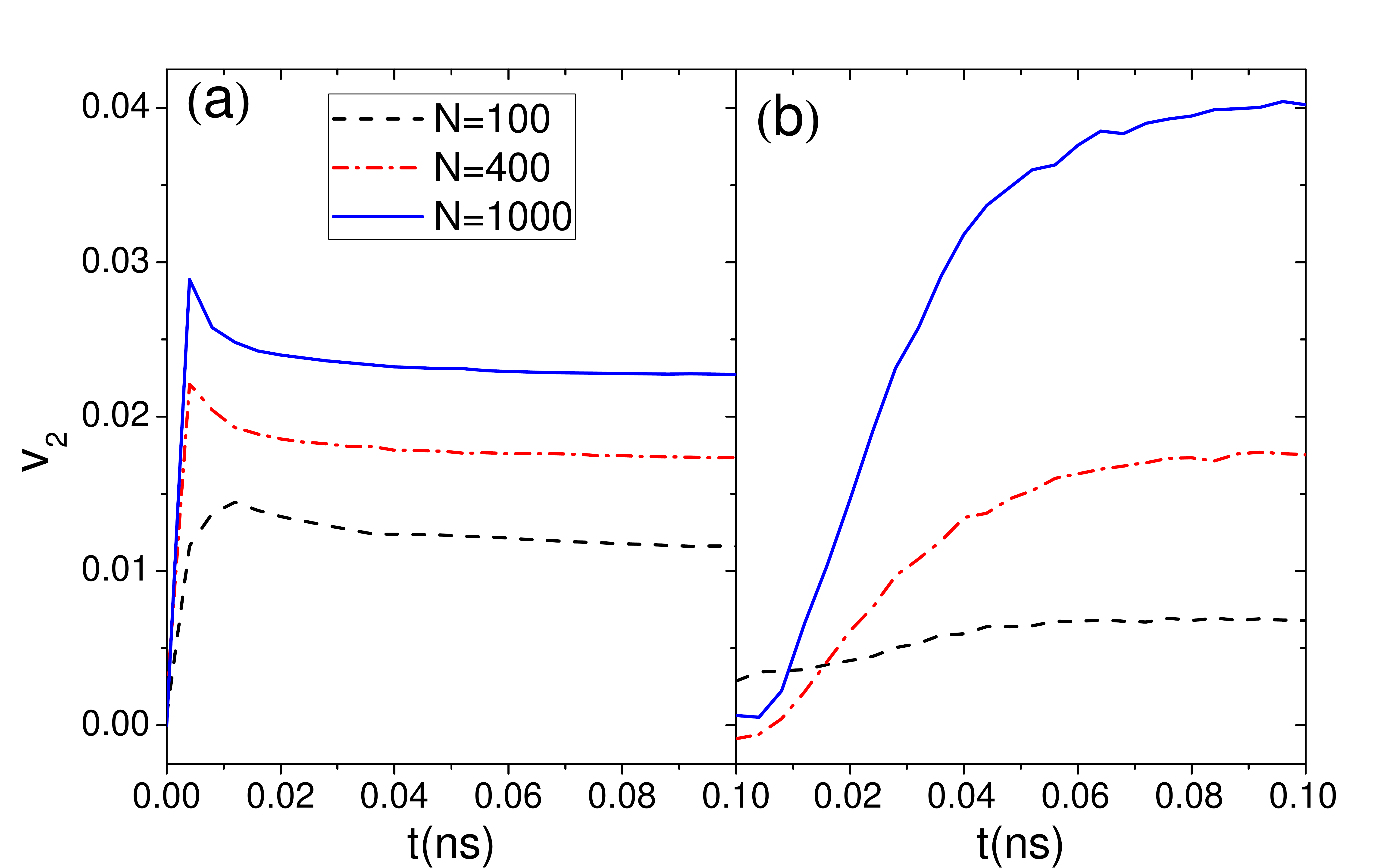}
\caption{(Color online) The elliptic flow parameter $v_2$ as functions of time, for (a) the Coulomb interaction, and (b) elastic scattering. Initial trap size: $a=10$~nm, $b=24$~nm, and $c=50$~nm. Three values for the number of particles $N$ are shown.}
\label{fig1}
\end{figure}

We first consider only the Coulomb interaction with the Argon ion system. At the initial moment, the ion system can be viewed as an ellipsoid of uniform charge distribution. The equipotential surfaces are elliptical. The electric field produced by the ions is stronger in the $x$ direction than in the $y$ direction, and the electric field on the periphery is larger than in the interior. The ions move under the influence of the elliptical electric field, gaining larger momentum in $x$ than in $y$ direction. This results in an elliptic flow in the ion system. This is shown by the finite $v_2$ in Fig.~\ref{fig1}(a) where three values of $N$ are depicted.

To compare with the long-range effect of the Coulomb interaction, we further simulate a dilute system of Argon atoms with only elastic scattering. After each scattering, the atom's motion is isotropized in the pair c.m.~frame whether the initial motion is in the $x$-direction or in the $y$-direction. However, because of the anisotropic population of the atoms in the $x$-$y$ space, atoms moving in the $y$ direction suffer more scatterings than those in the $x$ direction. More initially $y$-moving atoms are diverted into the $x$ direction than initially $x$-moving atoms being diverted into the $y$ direction. This results in fewer $y$-moving atoms than $x$-moving ones, hence a positive $v_{2}$ parameter. This is shown in Fig.~\ref{fig1}(b) where three values of $N$ are depicted.

\begin{figure}[hbt]
\centering
\includegraphics[width=0.75\textwidth]{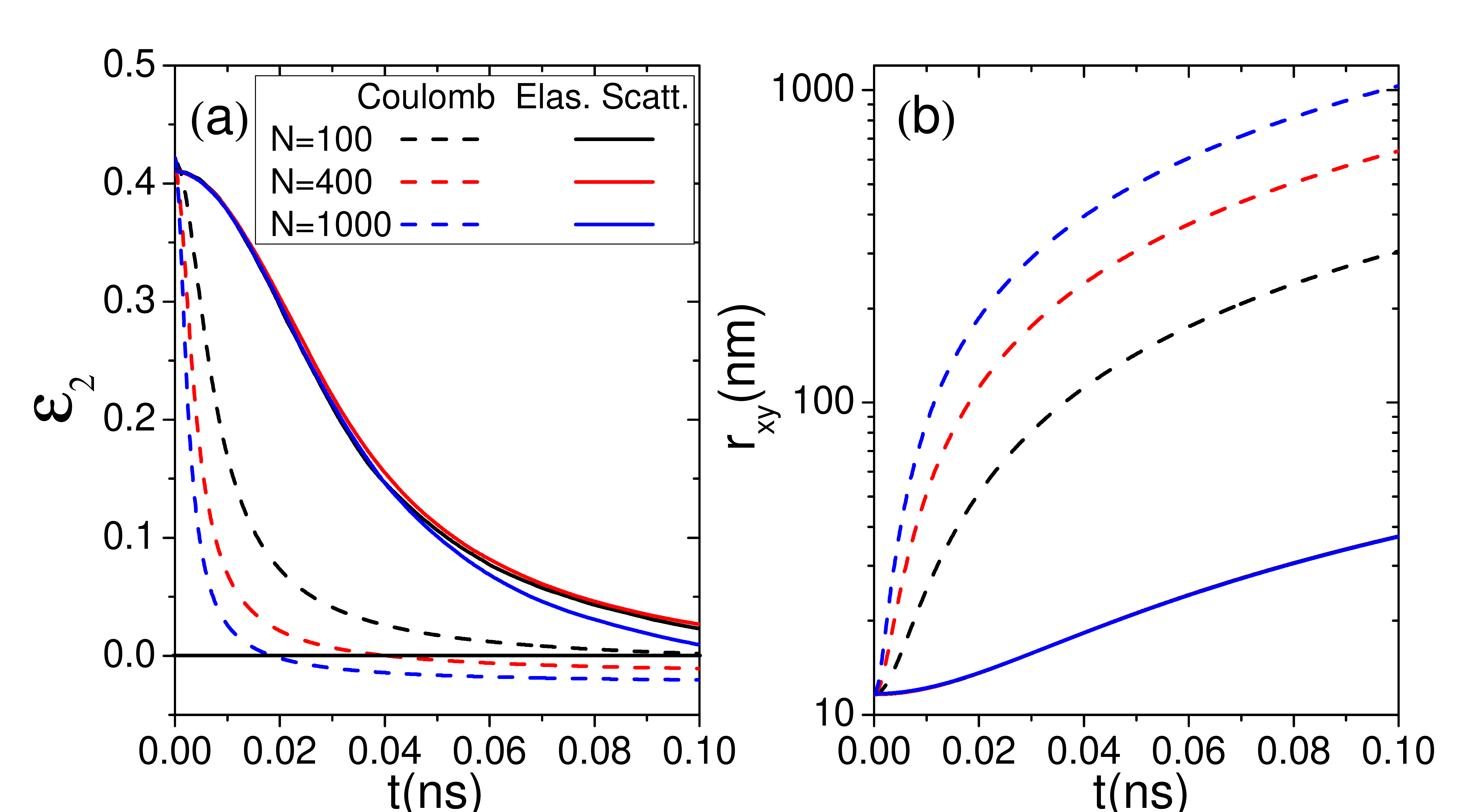}
\caption{(Color online) (a) The average $\varepsilon_{2}$ and (b) the average radial distance as functions of time. Initial trap size: $a=10$~nm, $b=24$~nm, and $c=50$~nm. Three values for the number of particles $N$ are shown.}
\label{fig2}
\end{figure}

The powerhouse for the momentum space anisotropy is the configuration space anisotropy of particle distribution, and the driving force to convert the configuration space anisotropy to momentum space anisotropy is the particle interaction, the Coulomb interaction or elastic scattering in our study. Therefore, the time evolution of $v_2$ must root in the spatial eccentricity of particle distribution. Figure~\ref{fig2}(a) shows the time evolution of the spatial eccentricity $\varepsilon_2$ for both cases. In the Coulomb interaction case, the electric potential energy is converted into the ions' kinetic energy. The ions gain momentum quickly (the ion velocity can quickly reach $10^{4}$~m/s); the ion system expands rapidly; see the dashed curve in Fig.~\ref{fig2}(b) where the average radial distance of the ions is shown as a function of time. The gain in velocity and the expansion rate are stronger for larger ion density. As a result, the spatial anisotropy diminishes quickly; $v_2$ quickly builds up and reaches maximum, as showed in Fig.~\ref{fig1}(a).

In the elastic scattering case, the atoms do not gain extra momentum on average. The initially random momentum directions become more radially aligned due to scatterings. The system expands slowly and the $\varepsilon_2$ decreases slowly; see the solid curves in Fig.~\ref{fig2}(a). Consequently, the $v_2$ builds up slowly and saturates when $\varepsilon_{2}$ diminishes after a relatively long time, as seen in Fig.~\ref{fig1}(b). No significant difference is observed in the time variations of $\varepsilon_{2}$ or radial positions of the atoms for the three cases of atom densities, shown in Fig.~\ref{fig2}(a) and ~\ref{fig2}(b), respectively. There is, however, a large difference in the $v_2$ at a given time. This can be understood as follows. During the same amount of time, the atoms move over the same distance on average because there is no gain in the atom velocity. Thus the $\varepsilon_{2}$ changes are similar during the same amount of time for the different densities. But, the denser the system, the larger scattering frequency. Because of the more scatterings during this time, more rearrangement of the momentum vector into $x$ direction is achieved, hence larger $v_2$ for the denser system.

Our simulations show that the very different Coulomb interaction and short-range contact interaction can both generate elliptic anisotropy. How the anisotropy develops differs, depending on details such as how quickly the spatial anisotropy diminishes. In particular, because of the rapid increase in ion's velocity and the rapid expansion of the ion system under the Coulomb interaction, the spatial eccentricity rapidly decreases and becomes slightly negative. This should contribute to the decreasing $v_2$ after it reaches maximum as shown in Fig.~\ref{fig1}(a). However, when $\varepsilon_2$ becomes negative, the system has grown very large. The negative $\varepsilon_{2}$ and the Coulomb interaction over the large volume are insufficient to cause the $v_2$ decrease observed in Fig.~\ref{fig1}(a). We believe the $v_2$ decrease is mainly due to the long-range nature of the Coulomb interaction, as follows. When two ions move in parallel in $x$ direction, their mutual Coulomb repulsion give them extra momentum kick in the $y$ direction, thus reducing their $v_2$ anisotropy.

The long-range Coulomb interaction and the short-range elastic scattering have an important distinction. The Coulomb interaction in our simulation may be considered as a classical field effect, having nothing to do with local thermal equilibrium or the existence of a pressure tensor. The flow buildup under the Coulomb interaction is therefore unlikely hydrodynamic. The anisotropy under the Coulomb interaction may more straightforwardly considered in the following way. The Coulomb equal-potential surfaces are elliptical, and hence force more particles towards to the large potential gradient (shorter axis) direction, resulting in positive $v_2$.

\begin{figure}[hbt]
\centering
\includegraphics[width=0.75\textwidth]{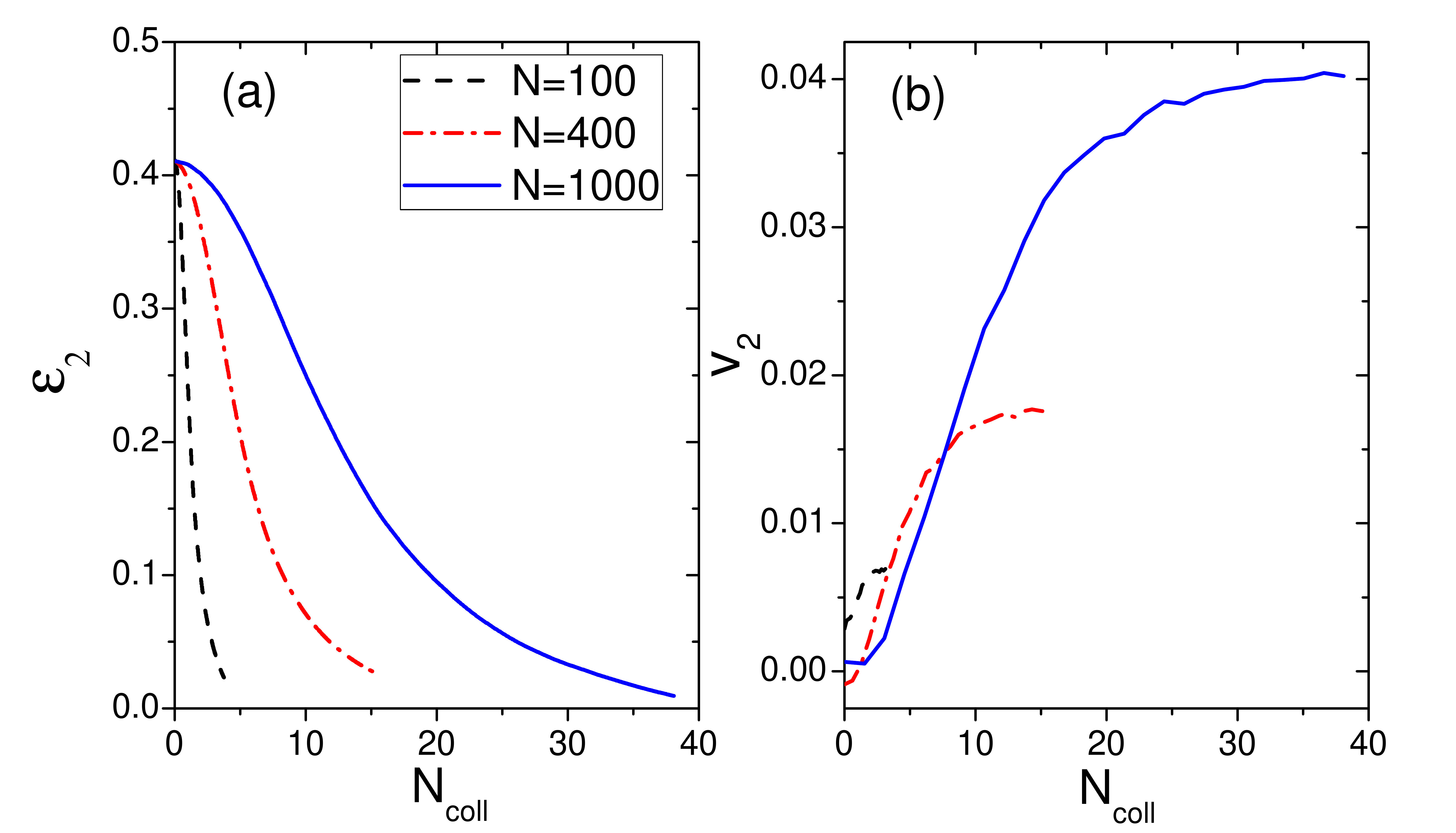}
\caption{(Color online) (a) The $\varepsilon_{2}$ and (b) $v_2$ after the given number of elastic scatterings, $\Ncoll$. Initial trap size: $a=10$~nm, $b=24$~nm, and $c=50$~nm. Three values of the number of particles $N$ are shown.}
\label{fig3}
\end{figure}

While the Coulomb interaction is continuous and cannot be viewed as individual collisions, elastic scatterings can be quantified by the number of collisions, $\Ncoll$. It is interesting to examine the development of $v_2$ as a function of $\Ncoll$, in addition to the time evolution studied in Fig.~\ref{fig1}(b). Figure~\ref{fig3} shows the $\varepsilon_2$ and $v_2$ as functions of $\Ncoll$ for three values of $N$. The $\varepsilon_2$ decreases sharply at initial stage. The $\varepsilon_2$ decreases to zero over different numbers of collisions. Because the time span between successive collisions is longer for smaller densities, it takes approximately the same amount of time to reach spatial isotropy (see Fig.~\ref{fig2}(a)). This makes sense because the average velocities are the same with different $N$, it takes approximately same amount of time to expand to the state of spherical symmetry. On the other hand, the $v_2$ follows roughly the same trend as a function of $\Ncoll$. This is because the $v_2$ is generated by interactions, and should depend on the amount of interactions. When $\varepsilon_2$ approaches to zero, the $v_2$ saturates as expected.

We have so far studied the time development of $v_2$. We now examine how the final-state integral $v_2$ depends on the two driving factors: the initial-state spatial eccentricity and the interaction strength. The initial eccentricity is varied by the trap aspect ratio in the $x$-$y$ plane. The interaction strength is varied by the particle density of the system.

\begin{figure}[hbt]
\centering
\includegraphics[width=0.75\textwidth]{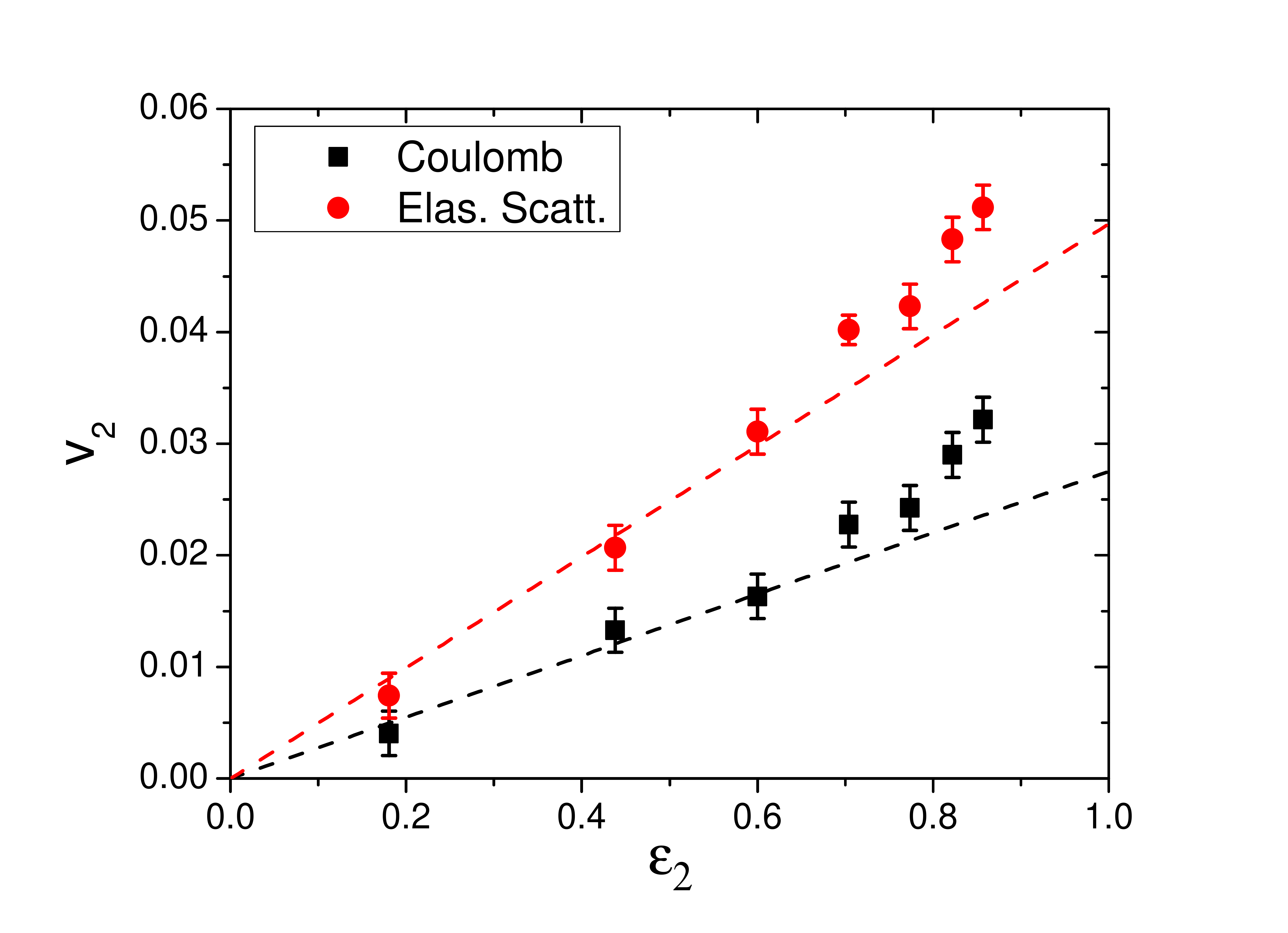}
\caption{(Color online) The final $v_2$ as a function of the initial $\varepsilon_{2}$. Initial trap size: $ab=240$~nm$^{2}$, $c=50$~nm. The number of particles is $N=1000$. The dashed lines are to guide the eye.}
\label{fig4}
\end{figure}

Figure~\ref{fig4} shows $v_{2}$ as a function of $\varepsilon_{2}$. We vary the initial trap $a:b$ aspect ratio, hence the $\varepsilon_2$, while keeping $c$ and the volume fixed. The $v_{2}$ increases linearly with $\varepsilon_{2}$ at small $\varepsilon_{2}$ (indicated by the dashed lines), after which the increase appears to be stronger than linear. This may indicate higher order effects in the $\varepsilon_{2}$ dependence of $v_2$.

\begin{figure}[hbt]
\centering
\includegraphics[width=0.75\textwidth]{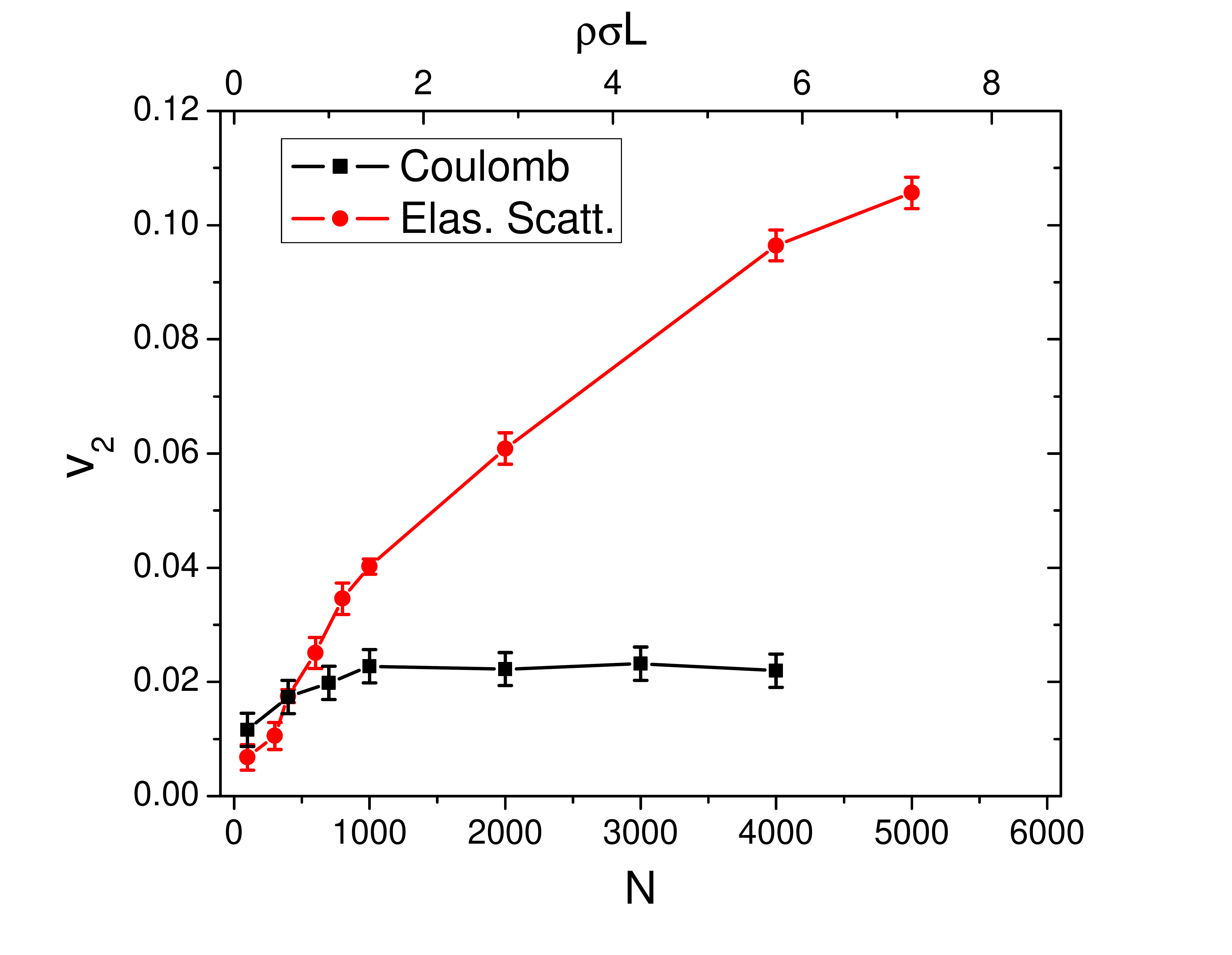}
\caption{(Color online) The final $v_2$ as a function of the number of particles $N$ for the two interaction cases: the Coulomb interaction and the elastic scattering. The values above the upper axis are the corresponding opacity $\op$ for the elastic scattering case only. Initial trap size: $a=10$~nm, $b=24$~nm, and $c=50$~nm.}
\label{fig5}
\end{figure}

Figure~\ref{fig5} shows $v_2$ as a function of $N$. The $v_2$ increases with $N$ for the elastic scattering case. The increase is less significant when $N$ becomes large. We can transform $N$ linearly into opacity $\op$ with the known cross section used in our simulation, $\op\equiv \frac{3}{4\pi}N\sigma(abc)^{-2}$; this is shown by the values above the upper axis in Fig.~\ref{fig5}. (Note these $\op$ values refer only to the elastic scattering case, not the Coulomb interaction case.) The $v_2$ is significant even for small $\op$, and increases rapidly with $\op$ at small $\op$ and less rapidly at large $\op$. The $v_2$ will presumably saturates when $\op$ approaches infinity.

In the case of the Coulomb interaction, the $v_2$ increases quickly at small $N$ and saturates for most of the simulated $N$ values. This may be expected because the Coulomb interaction is long range and its effective cross-section is infinite. The magnitude of $v_2$ is smaller for the Coulomb interaction case than the elastic scattering case for most of the simulated $N$. This is because the particles quickly gain speed under the Coulomb interaction so that the system quickly expands, diminishing the spatial eccentricity.

\section{Summary}
Anisotropic flow has been extensively studied in heavy ion collisions governed by the short-range strong interaction and QCD. Hydrodynamical and transport calculations involving quark and gluon degrees of freedom are the two main approaches. Hydrodynamical calculations assume high density and strong interactions. It is generally believed that strong interactions in a dense system are a necessary condition for the generation of anisotropic flow, converting the initial configuration space anisotropy into final-state momentum space anisotropy. However, the recent transport model studies by AMPT and MPC have casted this belief into doubt, suggesting that the anisotropic flow may be produced by the low density escape mechanism.

To shed light onto the situation, we have simulated the expansion of low density Argon ion and atom systems, initially trapped in an elliptical region. Significant elliptic anisotropies are found in the expansion under the Coulomb interaction and elastic scattering, respectively. The anisotropy increases with increasing initial spatial anisotropy and increasing particle density of the system (but still at low densities), to large values compatible to those from hydrodynamic calculations at high densities.
These results provide new insights and may help us understand the nature and the physics mechanisms of anisotropic flow in high energy heavy ion collisions and its role in the study of QCD, as follows.

This is the first simulation study of elliptic anisotropy of ion system under the Coulomb interaction. The study confirms that anisotropy development is universal, not only in the strong interaction, but also in Coulomb interaction.
The simulation of the short-range contact interaction of neutral atoms is performed by a home-made, straightforward two-body elastic scattering computer program. The anisotropy results qualitatively confirm those from AMPT and MPC with more extensive physics modeling.
The ion and atom systems we studied have rather low densities, but our results demonstrate that large elliptical flow develops in these low-density systems as well.
This suggests that large elliptic flow can not only develop in high density systems, as widely perceived, but also in low density systems. Large anisotropic flow is not unique to strongly interacting systems.

Given an observed large anisotropy, how does one tell if it is a result of a high-density strongly interacting (hydrodynamic) system or a low-density weakly interacting system? A unique distinction between the two is the development of strong collective radial flow in the former and a lack of strong radial flow in the later. We postpone the study of collective radial flow to a future work.

\section*{Acknowledgments}
We acknowledge the use of the computing server C3S2 in Huzhou University. This work was supported in part by the National Natural Science Foundation of China under Grants Nos.~11405054, 11375062, 11647306, and 11747312, and US~Department of Energy Grant No.~DE-SC0012910.

\bibliography{../ref}
\end{document}